\documentstyle[aps,prc,eqsecnum,epsfig]{revtex}
\newcommand{\bea}{\begin{eqnarray}}
\newcommand{\eea}{\end{eqnarray}}

\begin{document}
\draft
\tightenlines
%\preprint{
%\rightline{\vbox{\hbox{\rightline{McGill/99-28}}}}}

%\twocolumn[\hsize\textwidth\columnwidth\hsize\csname @twocolumnfalse\endcsname

\title{Hadronic interactions of the $J/\psi$}

\author{Kevin L. Haglin$^\dag$ and Charles Gale$^\ddagger$\\}

\address{ $\dagger$ Department of Physics, Astronomy and Engineering
Science\\St. Cloud State University, St. Cloud, MN 56301, USA\\}

\address{$^\ddagger$ Physics Department, McGill University\\
         3600 University St., Montr\'eal, QC, H3A 2T8, Canada}

\date{\today}

\maketitle
\begin{abstract}

We calculate the cross sections for reactions of the $J/\psi$ with
light mesons. We also evaluate its finite temperature spectral function.   
We investigate separately the role
of elastic and inelastic channels and we compare their respective
importance. We describe $J/\psi$ absorption channels that have not
been considered previously to our knowledge.  The relevance of our 
study to heavy ion collisions is discussed.
\end{abstract}

%\pacs{PACS: }

\section{Introduction }

The study of relativistic heavy ion collisions offers the tantalizing
possibility of observing many-body effects in a strongly interacting
system at densities and temperatures far removed from equilibrium. At
ultrarelativistic energies, the main focus of the active experimental
and theoretical programs is the creation, observation, and interpretation
of a new form of matter: the quark-gluon plasma. Its existence is a
prediction of QCD, even though some ambiguities concerning
the specific nature of an eventual phase transition and its 
experimental signatures \cite{qm99} still remain. It is fair to say
that the activity generated by this field makes it one of the most
exciting areas of contemporary subatomic physics. 

The $J/\psi$ meson has been singled-out as
a promising candidate to signal deconfinement. Indeed, 
the presence of a high temperature quark-gluon plasma would screen the
$c \bar{c}$ interaction \cite{MS} or ionize the quarkonium state 
\cite{dima_2}, leading to a suppression of the $J/\psi$ in
events where the plasma is produced in comparison with events where it
is not. While the suppression of $J/\psi$ (and $\psi^\prime$) in p-A and in
heavy ion collisions involving medium-mass projectiles at 200A GeV 
\cite{jpsix1} can be explained by absorption models without plasma
assumptions \cite{jpsit1,gjk}, the subsequent Pb + Pb data  has led to analyses
involving plasma formation \cite{dima}. However, the plasma
interpretation of this  ``anomalous''
$J/\psi$ absorption observed with the Pb projectile needs the
introduction of model-dependent assumptions. Furthermore, alternative
explanations which rest purely on hadronic grounds are starting to
appear \cite{hadth,cape2000}. Any model of $J/\psi$ suppression, whether it
includes hadronic ``comovers'' \cite{jpsit1} or not,  relies on
simple assumptions about the size of the cross sections with nucleons and
light hadrons. Unfortunately, up to recently only a few calculations of 
the interaction cross
section of $J/\psi$ with light hadrons could be found in the published
literature \cite{jpsixsect,jpsixsect2},
and the results of those calculations are not in agreement with each
other \cite{redlich}.  

Our aim in this work is the following. We plan to systematically
explore the different channels of interaction of the $J/\psi$ with
light hadrons and calculate the corresponding cross sections, in 
light of recent calculations with an effective hadronic Lagrangian 
\cite{kevin,linko}. We shall perform no attempts to find heavy ion data here.
However we will try to bring the study of the hadronic interactions of
the $J/\psi$ meson closer to the level of sophistication that the
lighter vector mesons ($\rho$, $\omega$, and $\phi$) currently enjoy. 
Consequently, we calculate the spectral function for the charmonium
state, in a gas of light mesons at finite temperature. Bear in mind
that this does not imply that the $J/\psi$ is thermalized. We view this
calculation as a necessary prelude to a more complete understanding of
the behavior of the charmonium bound states in hadronic matter at finite
temperature and density. Because of the direct decay into muon pairs, the
$J/\psi$ spectral function is directly accessible experimentally. 

Our paper is organized as follows: in the next section we discuss 
details of a heavy meson chiral Lagrangian bearing hadronic interactions
upon which our quantitative estimates are based. We also describe a
slight variant of this model that has been used in phenomenological
applications. The dominant channels in this study, both elastic and 
inelastic, are then considered and 
the associated cross sections are shown. We introduce new channels, to
our knowledge, for $J/\psi$ absorption on mesons. We then proceed to a 
discussion of the
scattering widths induced by the interactions. We will then show the
resulting $J/\psi$ spectral function, and explore its temperature and
momentum dependence. We summarize and conclude. 

\section{Chiral Lagrangian for light and heavy mesons}

We discuss here the basic assumptions and ingredients in our chiral Lagrangian
approach for light and heavy pseudoscalar and vector mesons.
We shall model the interaction of the charmonium state with 
lighter mesons through meson exchanges. In 
order to include charmed mesons, the smallest possible symmetry group
that potentially contains the relevant phenomenology is SU(4). However,
SU(4) is in fact badly broken by the large mass of the charmed quark.
This also can be seen in the poor agreement obtained between the
extended mass formula and the experimentally measured masses 
\cite{greiner}.    We adopt here the following pragmatic viewpoint: we
work here with the physical mass eigenstates and the physical mass
matrix will represent the relevant breaking of the original symmetry.
Furthermore, we compare two calibration methods, (i) the chiral
gauge coupling will be uniquely determined from light vector spectroscopy,
namely the $\rho$ meson, and (ii) relevant coupling constants
are individually chosen by either empirical constraints where they
exist or model calculations in the absence of measurement.
We feel that it is crucial for this effective approach to be in tune
with the largest possible range of phenomenology at the appropriate
energy scale.

Description of light and heavy pseudoscalars in a single framework
can be obtained using a four-flavor chiral Lagrangian.  The
basic nonlinear SU(4) $\sigma$ model apart from mass terms is
\begin{eqnarray}
{\cal L}_{0} & = & {-F_{\pi}^{2}\over 8}{\rm 
Tr}(\partial_{\mu}U\partial^{\mu}U^{\dag}),
\nonumber\\
U & = & {\rm exp}\left\lbrack{2i\phi\over F_{\pi}}\right\rbrack.
\end{eqnarray}
The constant $F_{\pi} \simeq$ 135 MeV and $\phi$ is
the pseudoscalar multiplet matrix.
Correct normalization leads to 
\begin{eqnarray}
\phi & = & \left(
\begin{array}{cccc}
\frac{\pi^0}{\sqrt{2}} + \frac{\eta}{\sqrt{6}} + \frac{\eta_c}{\sqrt{12}}
& \pi^+ & K^+ & \bar{D}^0 \\
\pi^- & -\frac{\pi^0}{\sqrt{2}} + \frac{\eta}{\sqrt{6}} +
\frac{\eta_c}{\sqrt{12}} & K^0 & D^- \\
K^- & \bar{K}^0 & - \eta \sqrt{\frac{2}{3}} + \frac{\eta_c}{\sqrt{12}}
& D_s^- \\
D^0 & D^+ & D_s^+ & - 3 \frac{\eta_c}{\sqrt{12}}  \\ 
\end{array}
\right) \ ,
\end{eqnarray}
To introduce vector mesons we make the replacement
\begin{eqnarray}
\partial_{\mu}U & \to & {\cal D\/}_{\mu}U \equiv \partial_{\mu}U
-igA_{\mu}^{L}U + igUA_{\mu}^{R},
\end{eqnarray}
and we add kinetic terms
\begin{eqnarray}
{\cal L}_{1} & = &   -{1\over 2}{\rm Tr}\left(
F_{\mu\nu}^{L}F^{L\,\mu\nu} +
F_{\mu\nu}^{R}F^{R\,\mu\nu}\right)
+  \gamma\,{\rm Tr}\left(F_{\mu\nu}^{L}UF^{R\,\mu\nu}U^{\dag}\right),
\end{eqnarray}
where $A_{\mu}^{L}$ and $A_{\mu}^{R}$ are the chiral spin-1 fields and where
\begin{eqnarray}
F_{\mu\nu}^{L} & = & \partial_{\mu}A_{\nu}^{L}-\partial_{\nu}A_{\mu}^{L}
-ig\left\lbrack A_{\mu}^{L}, A_{\nu}^{L}\right\rbrack,
\nonumber\\
F_{\mu\nu}^{R} & = & \partial_{\mu}A_{\nu}^{R}-\partial_{\nu}A_{\mu}^{R}
-ig\left\lbrack A_{\mu}^{R}, A_{\nu}^{R}\right\rbrack.
\end{eqnarray}
Next we add mass terms for the spin-1 fields
plus two generalized mass terms
\begin{eqnarray}
{\cal L\/}_{2} & = & -m_{0}^{2}\,{\rm Tr}\left(A_{\mu}^{L}A^{L\,\mu}  
+ A_{\mu}^{R}A^{R\,\mu}\right)
+ B\,{\rm Tr}\left(A_{\mu}^{L}UA^{R\, \mu}U^{\dag}\right)
+ C\,{\rm Tr}\left(A_{\mu}^{L}A^{R\, \mu} +A_{\mu}^{R}A^{L\, \mu}\right).
\end{eqnarray}
The aim for the present model is to describe the normal parity states, so
we must eliminate the axial-vector matrix field $A_{\mu} \equiv 
A_{\mu}^{L} - A_{\mu}^{R}$. To accomplish this, we follow the ideas 
presented in Ref.~{\cite{schecter85}} and make a gauge transformation resulting
in $A_{\mu}^{\prime} = 0$.  Equivalently, in the primed gauge,
$A_{\mu}^{L\, \prime} = A_{\mu}^{R\, \prime} \equiv \rho_{\mu}$.  
The vector meson matrix multiplet is
\begin{eqnarray}
\rho_{\mu}\ = \ \left(
\begin{array}{cccc}
\frac{\rho^0}{\sqrt{2}} + \frac{\omega}{\sqrt{6}} + \frac{J/\psi}{\sqrt{12}}
& \rho^+ & {K^*}^+ & \bar{{D^*}}^0 \\
\rho^- & -\frac{\rho^0}{\sqrt{2}} + \frac{\omega}{\sqrt{6}} +
\frac{J/\psi}{\sqrt{12}} & {K^*}^0 & {D^*}^- \\
{K^*}^- & \bar{{K^*}}^0 & - \omega \sqrt{\frac{2}{3}} + \frac{J/\psi}{\sqrt{12}}
& {D^*}_s^- \\
{D^*}^0 & {D^*}^+ & {D^*}_s^+ & - 3 \frac{J/\psi}{\sqrt{12}}  \\ 
\end{array}
\right)_{\mu} \ .
\end{eqnarray}
The specific choices $U^{1/2} = \xi$, $U^{-1/2} = \xi^{\dag}$ and
\begin{eqnarray}
A_{\mu}^{L} & = & \xi\rho_{\mu}\xi^{\dag} + {i\over g}\xi\partial_{\mu}
\xi^{\dag} \nonumber\\
A_{\mu}^{R} & = & \xi^{\dag}\rho_{\mu}\xi + {i\over g}\xi^{\dag}\partial_{\mu}
\xi \nonumber\\
U & = & \xi\mbox{\boldmath{$1$}}\xi
\end{eqnarray}
will ``gauge away'' the positive-parity states from the model by producing
the requisite $A_{\mu}^{\prime} = 0$.  

Gauging away the axial fields with the above mentioned
transformation yields the following Lagrangian (utilizing Hermiticity
$\phi$ = $\phi^{\dag}$ and $\rho_{\mu} = \rho_{\mu}^{\dag}$)
\begin{eqnarray}
{\cal L}_{0} & = & 0 
\nonumber\\
{\cal L}_{1} & = & \left(\gamma - 1\right){\rm Tr}\left[F_{\mu\nu}
(\rho)F^{\mu\nu}(\rho)\right],
\nonumber\\
{\cal L}_{2} & = & \left(B + 2C - 2m_{0}^{2}\right){\rm
Tr}\left(\rho_{\mu}\rho^{\mu}\right)
+ {2i(B - 2C - 2m_{0}^{2})\over g\,F_{\pi}^{2}}{\rm 
Tr}\left(\rho_{\mu}\left[\partial^{\mu}\phi,\phi\right]\right)
\nonumber\\
& \ & + {4C\over F_{\pi}^{2}}{\rm 
Tr}(\left[\phi,\rho^{\mu}\right]^{2})
- {(B + 2C + 2m_{0}^{2})\over g^{2}\,F_{\pi}^{2}}{\rm 
Tr}\left(\partial_{\mu}\phi
\partial^{\mu}\phi\right),
\label{ell012}
\end{eqnarray}
where we have defined
\begin{eqnarray}
F_{\mu\nu}(\rho) \equiv \partial_{\mu}\rho_{\nu}-\partial_{\nu}\rho_{\mu}
-ig\left[\rho_{\mu},\rho_{\nu}\right].
\end{eqnarray}
We notice that under the gauge transformation the original kinetic
piece for the pseudoscalars vanishes---it  reappears in ${\cal L\/}_{2}$.
${\cal L\/}_{1}$ becomes a kinetic Yang-Mills term
for the field $\rho_{\mu}$, while its correct normalization points to
$\gamma = $ 3/4.  Furthermore, ${\cal L\/}_{2}$ includes the mass terms for the
$\rho_{\mu}$ field, kinetic terms for $\phi$, as well, it includes
three-point and four-point interaction terms.  It is clear that
the mass term for $\rho_{\mu}$ spoils local gauge invariance.  In order
to leave ${\cal L\/}_{0} + {\cal L\/}_{1} + {\cal L\/}_{2}$
chirally and locally gauge invariant, we choose $B +2C - 2m_{0}^{2} = 0$. 
Omitting kinetic energy terms, we arrive at the
model's chiral and gauge invariant set of interactions.  They
are the following
\begin{eqnarray}
{\cal L\/}_{\rm\, int} & = & ig\,{\rm Tr}
\left(\rho_{\mu}\left[\partial^{\mu}\phi,\phi\right]\right)
- {g^{2}\over 2}{\rm Tr}(\left[\phi,\rho^{\mu}\right]^{2})
+ ig\,{\rm Tr}\left(\partial_{\mu}\rho_{\nu}\left[\rho^{\mu},\rho^{\nu}
\right]\right)
+ {g^{2}\over 4}{\rm Tr}(\left[\rho^{\mu},\rho^{\nu}\right]^{2})
\label{ellint}
\end{eqnarray}
where, for convenience, we have attached physical significance to
the gauge coupling constant through 
\begin{eqnarray}
{2(B - 2C -2m_{0}^{2})\over g\,F_{\pi}^{2}} & = &
{g_{\rho\pi\pi}\over 2} \ = \ g.
\end{eqnarray}

Before leaving the formalism we summarize up to this point.  An effective 
chiral Lagrangian of light plus heavy pseudoscalar and vector mesons 
has been constructed to be fully chirally U(4)$\times$U(4) invariant. 
In particular, no loose ends are left in the model since the axial
fields have been completely gauged away. The 
result is written in Eq.~(\ref{ell012}).  From here we
further impose local gauge invariance
and arrive at interaction terms given in Eq.~(\ref{ellint}).
The model has two input parameters: $F_{\pi}$ and $g_{\rho\pi\pi}$.
Since we adjust $g_{\rho\pi\pi}$ using the decay rate into pion pairs
at the physical rho pole, we implicitly also use $m_{\rho}$.
The matrix algebra implied in the above expressions can now be 
explicitly carried out to obtain Lagrangians involving specific physical 
fields.  

\subsection{Chiral Model Predictions}
 
The model is constructed in some sense to pivot off the rho meson since
$\rho$ decay into two pions uniquely determines the gauge coupling 
constant $g$.  Taking $\Gamma_\rho$ = 151 MeV, and $m_{\rho}$ = 770 MeV, 
we find $g_{\rho\pi\pi} = {\rm 2}g \simeq$ 8.54.  Having fixed the single
parameter in Eq.~(\ref{ellint}), we are in position to ``predict'' 
widths for meson decays where phase space is open, and we are particularly
interested in the strangeness and charm sectors.  $K^{*}$'s and $D^{*}$'s 
are chosen to test the 
symmetry breaking effects in the extremes.   Results for widths are 
listed in Table \ref{table1}.  We find $K^{*}$ widths consistent to
within 10\% of experiment, and $D^{*}$ widths consistent with
other model calculations\cite{col94,rob99,nav00}.    We are
hesitant to read too much into these numbers, but they 
begin to suggest that the symmetry breaking effects might
largely be accounted for merely by using the physical mass eigenstates.

All coupling strengths are now in principle fixed.
We have first evaluated absorption cross sections for reactions involving 
the $J/\psi$ in the initial state.  We list the processes under 
consideration in Table \ref{table2}. Note that
each process can in principle involve several Feynman diagrams and their
interference. 

\begin{figure}[htbp]
  \begin{center}
  \epsfxsize 80mm 
  \epsfbox{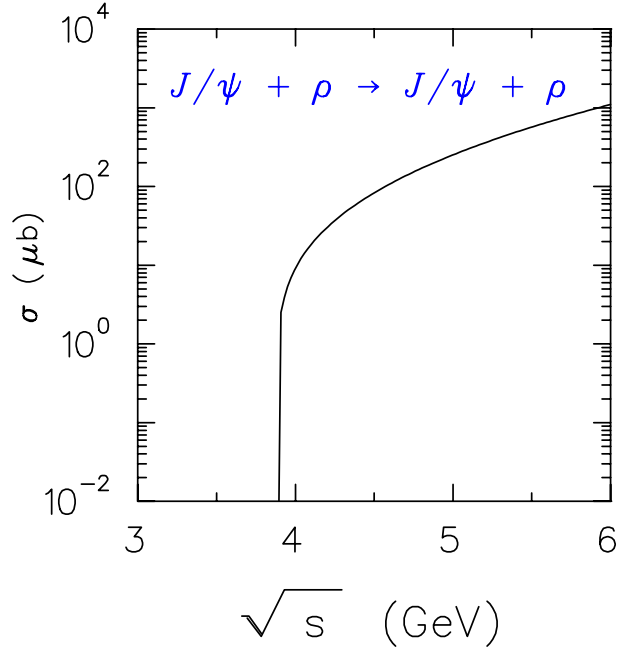}
    \caption{Elastic cross section {\it J\/}/$\psi$ + $\rho
             \rightarrow$ {\it J\/}/$\psi$ + $\rho$.
            }
    \label{fig1}
  \end{center}
\end{figure}

For the sake of brevity, we do not show all the cross sections we 
have evaluated. We shall
restrict ourselves here to the most important ones. Our findings first support
the notion that the elastic channels are quantitatively small. Apart
from the $\omega$ to be discussed later, the largest
contribution in this category is shown in Fig. \ref{fig1}, and is $J/\psi +
\rho \rightarrow J/\psi + \rho$. By $\sqrt{s}$ = 6 GeV, the cross section
has just about risen to 1 mb.  This process is modeled by pion exchange
and therefore involves 
vector-vector-pseudoscalar interactions which are not included in the 
chiral Lagrangian.  Instead, the relevant Lagrangian is
\begin{eqnarray}
{\cal L} & = & g\, \epsilon_{\alpha \beta \mu \nu}\, \partial^\alpha V^\beta
\partial^\mu V^\nu\, \phi\ .
\label{lvvp}
\end{eqnarray}
The coupling strength is fixed to match the  measured \cite{pdg} 
$J/\psi\to\pi\rho$ decay width.
The elastic process immediately following this one in importance is $J/\psi
+ \eta \rightarrow J/\psi + \eta$, and the corresponding cross section is
only 6 nb at the same value of $\sqrt{s}$. All other elastic processes
listed in Table \ref{table2} are at the nb, pb, and even fb level.

\begin{figure}[htbp]
  \begin{center}
  \epsfxsize 80mm 
  \epsfbox{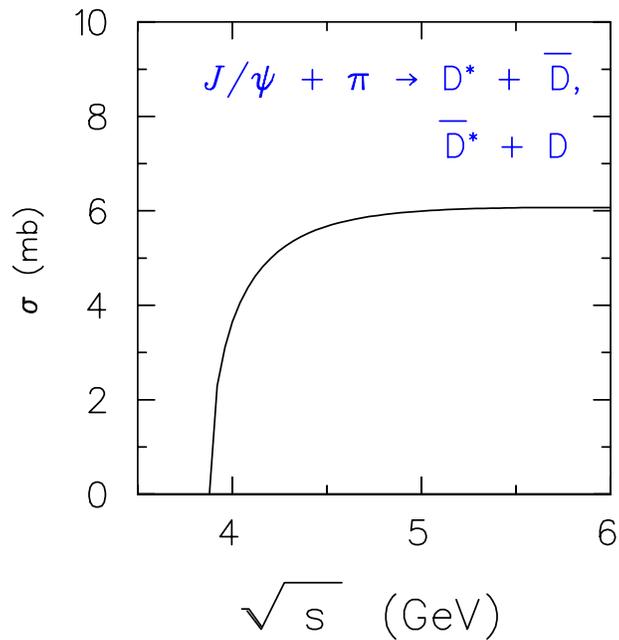}
    \caption{Isospin averaged total cross section for {\it J\/}/$\psi$ + $\pi
             \rightarrow$ $(\bar{\it D\/}$ + {\it D\/}$^{*})$
              + $(\bar{\it D\/}^{*}$ + {\it D\/}).
            }
    \label{fig2}
  \end{center}
\end{figure}

\begin{figure}[htbp]
  \begin{center}
  \epsfxsize 80mm 
  \epsfbox{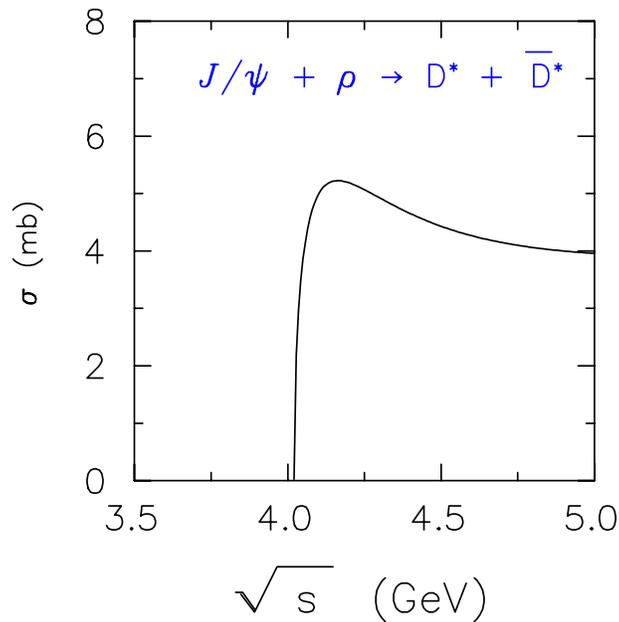}
    \caption{Cross section for {\it J\/}/$\psi$ + $\rho
             \rightarrow$ $\bar{\it D\/}^{*}$ + {\it D\/}$^{*}$.
            }
    \label{fig3}
  \end{center}
\end{figure}

\begin{figure}[htbp]
  \begin{center}
  \epsfxsize 80mm 
  \epsfbox{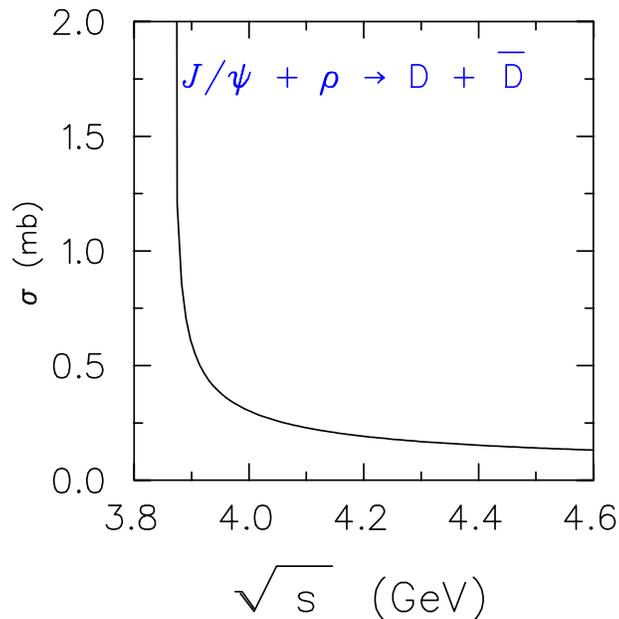}
    \caption{Cross section for {\it J\/}/$\psi$ + $\rho
             \rightarrow$ $\bar{\it D\/}$ + {\it D\/}.
            }
    \label{fig4}
  \end{center}
\end{figure}

We now turn to the most important channels in our study: inelastic
reactions. We show in Fig. \ref{fig2} the isospin averaged total
absorption cross section $J/\psi + \pi \rightarrow D^* + \bar{D}$ plus
$\bar{D}^{*} + D$. 
Next in importance is the reaction $J/\psi + \rho \rightarrow D^*
\bar{D^*}$, this is shown in Fig. \ref{fig3}.
Note that exothermic reactions are also possible, but they typically settle
down to a low cross section value. A representative example is $J/\psi +
\rho \rightarrow D + \bar{D}$, and this is shown in Fig. \ref{fig4}.

\subsection{Phenomenological Model}

To compare with other calculations which have used effective Lagrangian
methods but have constrained the model differently, we include this
subsection.  If we retreat somewhat from the symmetry and allow the
coupling constants to be separately adjusted to empirical constraints
or models we arrive at different predictions.  For instance, if we
choose the $D^{*}D\pi$ coupling constant to give a width consistent
with a relativistic potential model prediction of 46 keV for
$D^{*\, \pm}$\cite{Co94}, we find $g_{D^{*}D\pi}$ = 4.4 (whereas, the chiral
prediction from the previous subsection used a value 3.02).
Vector dominance arguments have been further used to fix couplings
like $g_{J/\psi\,DD}$ and $g_{J\/\psi\,D^{*}D^{*}}$ to 
be 7.7\cite{kevin,linko}. 
Again, the chiral prediction used above is 4.93.  We stress
that the chiral model
\begin{figure}[htbp]
  \begin{center}
  \epsfxsize 80mm 
  \epsfbox{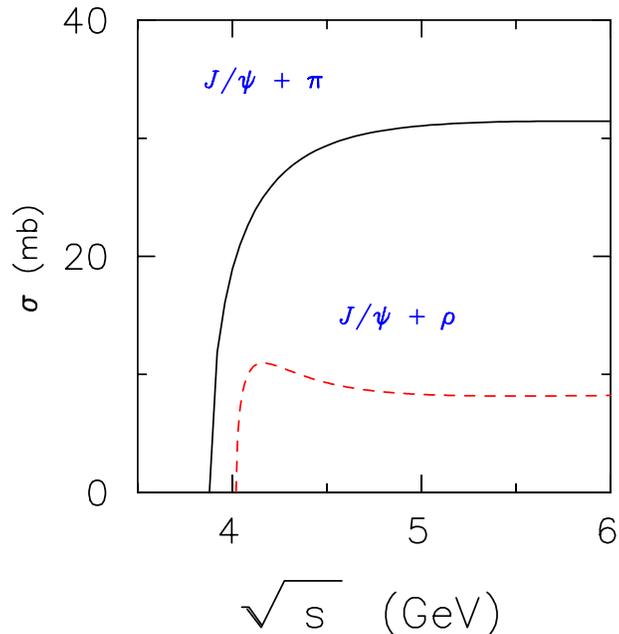}
    \caption{Cross sections for pion- and rho meson-induced dissociation
             of $J/\psi$.  The specific channels are the same as
             Figs.~\protect{\ref{fig2}} and \protect{\ref{fig3}}.
            }
    \label{fig4b}
  \end{center}
\end{figure}
calculations are not different from the previous effective Lagrangian
methods, practically speaking, the differences are merely
methods of calibration \cite{expln}. 
Note however that the calibration method we associated with the
``phenomenological model'' will lead
to K$^*$ phenomenology that is off by a factor of two, as seen in
Table~\ref{table1}.

In Fig. \ref{fig4b} we show pion and rho dissociation of $J/\psi$
in the phenomenological approach.  The results are to be compared
with the chiral model predictions in Figs.~\ref{fig2} and \ref{fig3}.
 So in some sense, the different results
could be viewed as representative of uncertainties in the present hadronic
approaches to $J/\psi$ dynamics. In this work the preference will go
to the so-called chiral approach as it generates the hadronic phenomenology
contained in Table~\ref{table1} which does not contradict experimental
measurements. 

\section{Anomalous processes}

We include next a section which reports on a couple of processes
in the anomalous sector which turn out to give significant
cross sections.  Based on the observation that $J/\psi\to\eta_{c} + \gamma$
is 1.3\% of 87 keV, we use vector dominance to estimate the coupling
of $J/\psi$ to $\eta_{c}$ and $\omega$.  The Lagrangian we use is
again listed in Eq.~(\ref{lvvp}).  We find a value $g_{J/\psi\eta_{c}\omega}$
= 9.5 GeV$^{-1}$.  Using similar reasoning and calculations for
$\omega\to\pi^{0}\gamma$, we extract $g_{\omega\rho\pi}$ = 11.6 GeV$^{-1}$.
Equipped with these couplings, we can compute the cross section
for $J/\psi + \pi \to \eta_{c} + \rho$ through $\omega$ exchange.  We 
present it in 
\begin{figure}[htbp]
  \begin{center}
  \epsfxsize 80mm 
  \epsfbox{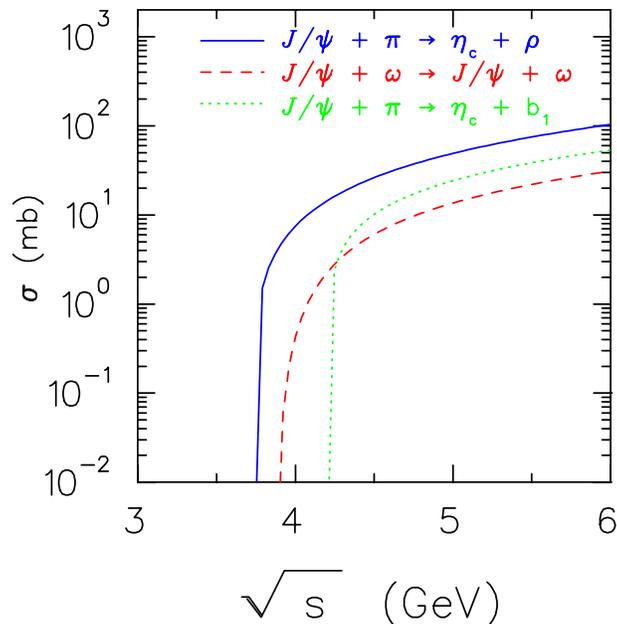}
    \caption{Cross sections for $J/\psi$ involving anomalous couplings and
             the $\eta_{c}$ meson. The specific channels are discussed in the
             text.
            }
    \label{anomalous}
  \end{center}
\end{figure}
Fig.~\ref{anomalous}, and remark that it is quite large. 
Since this calculaton has been done, we have found that Shuryak and
Teaney  had
considered this process previously using a model different from what is
done here \cite{shte}.

Similarly, the cross section  for $J/\psi + \pi \to \eta_c + b_1$ is estimated.
The value for the coupling $g_{b_1 \pi \omega}$ is deduced from the
measured decay width of the $b_1$ \cite{pdg}. The cross section is
displayed in Fig.~\ref{anomalous}. Its value is also large.
The elastic channel
$J/\psi + \omega \to J/\psi + \omega$ proceeding through through $\eta_{c}$
exchange can be considered.  It too is included in 
Fig.~\ref{anomalous}, and one can also
conclude that it is relatively important. We have also calculated
several other new reactions which we report on in the next section.

With these cross sections, and the ones associated with all the other
processes we have considered, the  
$J/\psi$ spectral function in a finite temperature gas of mesons can
now be calculated. However
before moving on to that topic, the important issue
of form factors needs to be addressed. 

\section{Hadronic form factors}

The field theory in this work is formulated in a hadronic language 
and does not 
deal with fundamental fields but with
degrees of freedom that are composite in terms of quark content. It is
clear that the exchange of heavy mesons (as the open charm $D$ mesons,
for example) leads to an interaction too short-ranged for the
interacting particles to be left unmodified \cite{malisg}. 
Meson-exchanges are perhaps parameterizations of other
phenomena which should be more evident at the parton level.  This
fact reveals itself in the appearance of hadronic form factors at
the interaction vertices of Feynman diagrams.
Those can be
the source of additional uncertainty in the model. 
Note that form factor considerations are not restricted to meson
exchange models like the one discussed here. For example, they occur 
also in the flux tube breaking model \cite{flux} and the ${^3}P_0$ model
\cite{3p0}. The
choice of form factors is
guided here by physical arguments, and they are introduced 
in a way that respects gauge invariance in the electromagnetic sector,
and Lorentz symmetry. 

A generic form is first chosen to construct $t-$ and $u-$channel 
hadronic form factors. A candidate
that lends itself to practical calculations is the monopole:
\bea
F(t) = { \Lambda^2 \over \Lambda^2 + |\, t-m_{\alpha}^{2}| }\ ,
\label{ff}
\eea
where $m_\alpha$ is the mass of the exchanged meson.
An advantage of this functional form is that the form factor
is normalized to 1 for on-shell particles. Also, even if the
kinematics venture into regions of time-like momentum transfer, this
choice of form factors remains unitary. Each vertex therefore receives
a contribution $F(t)$ or $F(u)$, depending on the appropriate 
kinematics. The vertices for the tadpole diagrams are determined by
replacing its metric tensor structure by a general tensorial
expansion constructed from the metric tensor and the available four-vectors.
The coefficients of this expansion are then chosen such that the total
amplitude is gauge invariant in the electromagnetic sector. The form
factors in this work therefore build Feynman
amplitudes that are both Lorentz- and gauge-invariant.

Our values for $\Lambda$ stem from elements of hadronic phenomenology 
which we describe now.  We first 
fix the coupling constant $g_{J/\psi \rho \pi}$ by reproducing the
measured width $\Gamma_{J/\psi \to \rho \pi}$. 
The appropriate vector-vector-pseudoscalar ($V V \phi$) Lagrangian has
been shown in Eq.~(\ref{lvvp}).
For the determination of $g$ the form factor (Eq.~\ref{ff}) plays no
role, by construction. $\Lambda$ is then determined by pushing one of the
particles off-shell. Consider the measured \cite{pdg} radiative decay 
width $\Gamma_{J/\psi \to \gamma \pi^0}$. Using a Vector Meson
Dominance (VMD) argument, one may assume that the photon couples to the
$\rho$ of the above strong interaction vertex. Then, with $g$ being
determined, $\Lambda$ can be obtained from a fit to the radiative decay
width.  A word of caution is necessary
here: the $J/\psi$ also has a $G$-parity violating decay like $J/\psi
\to \omega \pi$, so that presumably the photon could also originate from
the $\omega$ through VMD. However when compared with $J/\psi \to
\rho^0 \pi^0$, the decay into $\omega \pi^0$
is suppressed by an order of magnitude so that we can safely ignore it
here. In order to reproduce the $J/\psi$ radiative decay width one
needs the parameter in Eq.~\ref{ff} to be 
$\Lambda$ = 1.25 GeV.  This number is satisfying as it does represent a
scale that is typically associated with soft hadronic interaction as are
commonplace in, for example, the Bonn potential \cite{bonn}. 

Another method to pin down hadronic form factors consists of
considering $J/\psi$ and open charm photoproduction data and to use 
their relation  with $J/\psi$-nucleon total elastic and inelastic cross
sections \cite{bauer}. Using this argument, a $J/\psi$ + $N$ inelastic
cross section can be extracted from the data, and its value is 
$\approx$0.1 mb at $\sqrt{s}$ = 6 GeV \cite{rsz}. Below this energy,
some doubts have been expressed on the reliability of the cross section
extraction through VMD \cite{rsz}. We estimate the largest contribution
to the $J/\psi$-$N$ inelastic cross section to be $J/\psi N \to
\Lambda_c \bar{D}$. Using the form factor described above and requiring
that the inelastic cross section be 0.1 mb at $\sqrt{s}$ = 6 GeV sets
$\Lambda$ = 3 GeV. Past this energy value the $J/\psi$-$N$
meson-exchange cross section drops (unlike what is shown in Fig. (5) of
Ref. \cite{rsz}), so that the upper bound set by photoproduction data is not
exceeded. 
%Another possibly important reaction for nuclear $J/\psi$ 
%absorption is $J/\psi  \to       $. Some doubts on the importance of
%this specific channel stem from the uncertainty in the strong meson
%exchange coupling constant $g$.
With this in mind $\Lambda$ = 2 GeV is set as a conservative upper
bound for the remainder of this work, thereby allowing for the
contribution of other channels to the $J/\psi - N$ cross section. 

\begin{figure}[htbp]
  \begin{center}
  \epsfxsize 80mm 
  \epsfbox{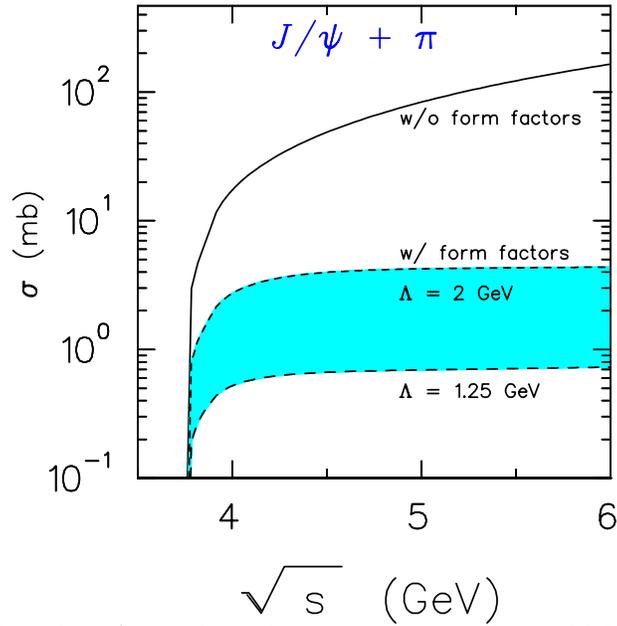}
    \caption{The value of the total inclusive $J/\psi$ + $\pi$ 
inelastic cross section,
and its sensitivity to our choices of the form factor parameter  
$\Lambda$.  }
    \label{difflambd}
  \end{center}
\end{figure}

\begin{figure}[htbp]
  \begin{center}
  \epsfxsize 80mm 
  \epsfbox{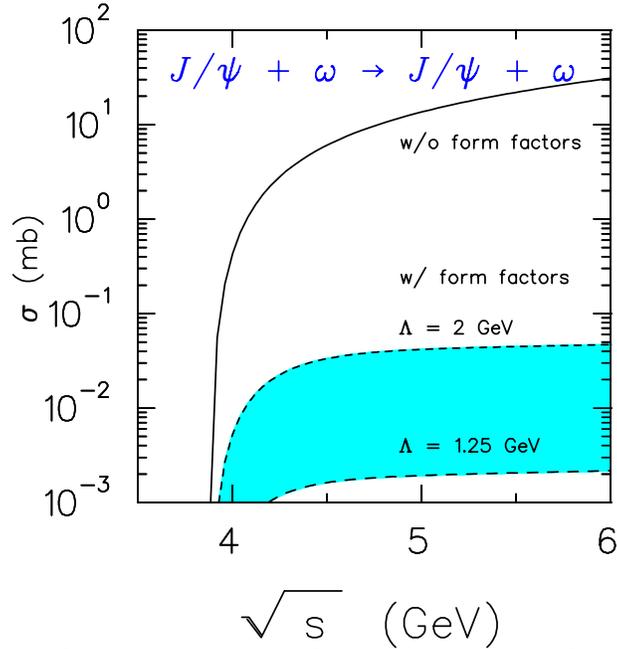}
    \caption{The value of the $J/\psi$ + $\omega$ elastic cross section,
and its sensitivity to our choices of the form factor parameter  
$\Lambda$. The solid curve is obtained without form factors, the dashed
curves mark the limits of our range in form factors.}
    \label{om_elas_wff}
  \end{center}
\end{figure}

Summarizing, a
range in $\Lambda$ was estimated for the hadronic form factor 
introduced in the
meson exchange model. Guided by hadronic phenomenology, we set 
1.25 GeV $\leq \Lambda \leq$ 2 GeV. 
The final cross sections are quite sensitive to
the choice of the cutoff parameter $\Lambda$. This sensitivity is 
shown in Fig.~\ref{difflambd} using the total inclusive absorption 
cross section
for $J/\psi$ on $\pi$. This cross section is the sum of the ones in
Figs.~\ref{fig2} and \ref{anomalous}. Even within the window that has
been determined for $\Lambda$, the cross section remains uncertain
within an order of magnitude. The larger value of the form factor
parameter has the total absorption cross section flattening out at
around 4 mb. This value is dominated by the cross section in the
anomalous sector with an $\eta_c$ in the final state, and is thus quite
insensitive to the choice of the chiral model or the phenomenological model 
to calculate $J/\psi + \pi \to D^* + \bar{D}$ + h.c. The apparent kink
in the low energy region of Fig.~\ref{difflambd} is related to the
different thresholds for the reactions in Fig.~\ref{fig2} and
\ref{anomalous}.  
Also note that Shuryak and Teaney estimate that the cross section for
$J/\psi + \pi \to \eta_c + \rho$ is 1.2 mb, using non-relativistic
quark model arguments \cite{shte}.  The value
obtained in the approach followed in this work is $\approx$ 1 mb in
the middle of the form factor range defined above. Those two
different methods thus give numerical results that are not inconsistent 
with each other. It is important in this nonperturbative sector to
cross-check model calculations. In this respect, the form factor
corrected cross section for the reaction of Fig.~\ref{fig2} has a mean
value of $\approx$ 0.1 mb, only slightly lower than that obtained in a
quark-interchange model \cite{wong}. 
The elastic cross section of the $J/\psi$ with the $\omega$ which was
shown in Fig.~\ref{anomalous} is also drastically affected by form factor
considerations. This is displayed in Fig.~\ref{om_elas_wff}. Note that
the suppression due to the form factor is different in the elastic and
inelastic cases, owing to different kinematics. 
\begin{figure}[htbp]
  \begin{center}
  \epsfxsize 80mm 
  \epsfbox{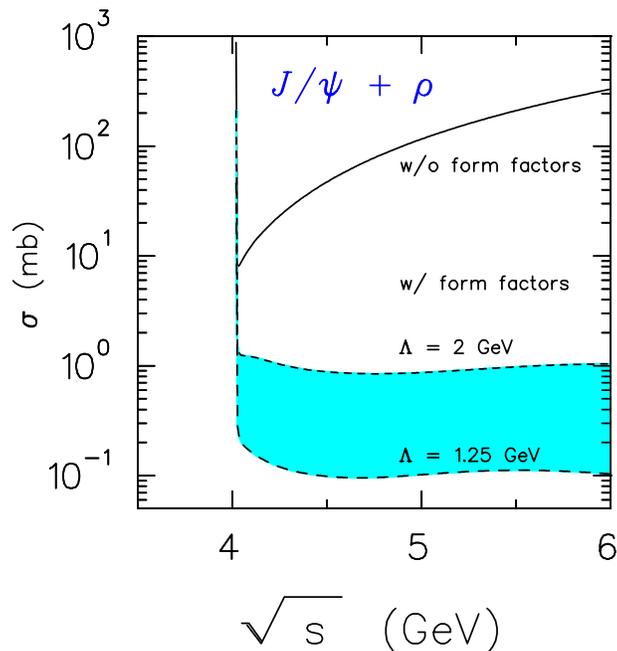}
    \caption{The value of the total inclusive $J/\psi$ + $\rho$ 
inelastic cross section,
and its sensitivity to our choices of the form factor parameter  
$\Lambda$. }
    \label{jpsi_rho_tot}
  \end{center}
\end{figure}
Finally we also show the total inclusive cross section for 
rho-induced absorption
of the $J/\psi$ in Fig.~\ref{jpsi_rho_tot}. It is worth noting 
that the upper choice of our form
factor parameter has a large energy limit of $\approx$ 1 mb.
Using the methods just outlined, we have estimated also $J/\psi + \eta
\to \eta_c + \phi$ (0.06 mb), $J/\psi + K \to \eta_c + K_1$ (0.18 mb).
The numbers in parentheses refer to cross section values that are
approximately in the centre of the form factor window.   
To our knowledge, those channels also have not been discussed previously.

\section{The $\mbox{\boldmath{$J$}}$/$\mbox{\boldmath{$\psi$}}$ spectral
function}

For hadrons immersed in a strongly interacting medium, one finds that
two-loop effects dominate over one-loop effects in the calculations of
imaginary parts of particle propagators \cite{kevwidth}. This owes
largely to the size of the coupling in the confined sector.  We have
verified that this is the case here also, by doing explicit calculations. Thus,
we neglect one-loop effects. We will also neglect effects on the real part
of the $J/\psi$ propagator. We thus will assume that the $J/\psi$ will
suffer negligible mass-shifts.  We partly base this reasoning on the 
large mass
of the charmonium vector meson. Also, recent calculations of 
$J/\psi$ properties in
nuclear matter do yield mass shifts that are small \cite{jpsishift}. 

Our first task is then to calculate the broadening due to collisions of
the $J/\psi$ with particles that make up the heat bath. The width
induced by a reaction of the type $J/\psi \, 2 \rightarrow 3 4$, where 2,
3, and 4 are arbitrary species is \cite{kevwidth,collwidth}. 

\bea
\Gamma (\omega, \vec{p} )\ = \ \frac{1}{2 \omega} \int \, d\Omega\, n_2 (E_2) (1 + n_3
(E_3)) (1 + n_4(E_4))| \overline{ {\cal M} (J/\psi 2 \rightarrow 3 4)}|^2\ ,
\label{width}
\eea
where $\omega\ = \ \sqrt{\vec{p\,}^2 + m_{J/\psi}^2}$, $\vec{p}$ being
the three-vector of the $J/\psi$.  Note that 3 or 4 can be a $J/\psi$. 
In Eq. (\ref{width}), 
\bea
d\Omega\ = \ d\bar{p}_2 d\bar{p}_3 d\bar{p}_4 (2 \pi)^4 \delta ( p + p_2
- p_3 - p_4 ) ,
\eea
and
\bea
d \bar{p}_i\ = \ \frac{d^3 p_i}{(2 \pi )^3\, 2 E_i}\ .
\eea 
The reactions we consider will involve only mesons.  In principle, a
$J/\psi$ can also be produced by an inverse reaction involving
particles from the thermal background \cite{collwidth}. Here, phase
space considerations make the inverse channel negligible. 

We write the spectral function of the $J/\psi$ as
\bea
A_{J/\psi} (\omega, \vec{p\,})\ = \ - 2 {\rm Im} D_{J/\psi} (\omega,
\vec{p\,})\ ,
\eea
where $D_{J/\psi}$ is the scalar part of the $J/\psi$ propagator.
Neglecting the difference between longitudinal and transverse
polarizations of the $J/\psi$ in the finite temperature medium
\cite{gk}, one has
\bea
D_{J/\psi} (\omega, \vec{p\,})\ = \ \frac{1}
{p^2 - m_{J/\psi}^2 - F (\omega, \vec{p\,})}\ ,
\eea
where $p^\mu =  (\omega, \vec{p\,} )$, and $F$ is the scalar 
imaginary self-energy.  Then, using 
\bea
\Gamma_{J/\psi}=- 1 / m_{J/\psi}~{\rm Im}~F({p^2}=m_{J/\psi}^2)
\eea
one can write in the on-shell approximation
\bea
A_{J/\psi} (\omega, \vec{p\,})\ = \ \frac{2\, m_{J/\psi}\, 
\Gamma_{J/\psi}}{(p^2 - m_{J/\psi}^2)^2 + m_{J/\psi}^2
\Gamma_{J/\psi}^2 }\ ,
\eea
where $\Gamma_{J/\psi}$ contains all the contributions we have
discussed so far: the vacuum width and the contributions from elastic
and inelastic collisions. 

\begin{figure}[htbp]
  \begin{center}
  \epsfxsize 80mm 
  \epsfbox{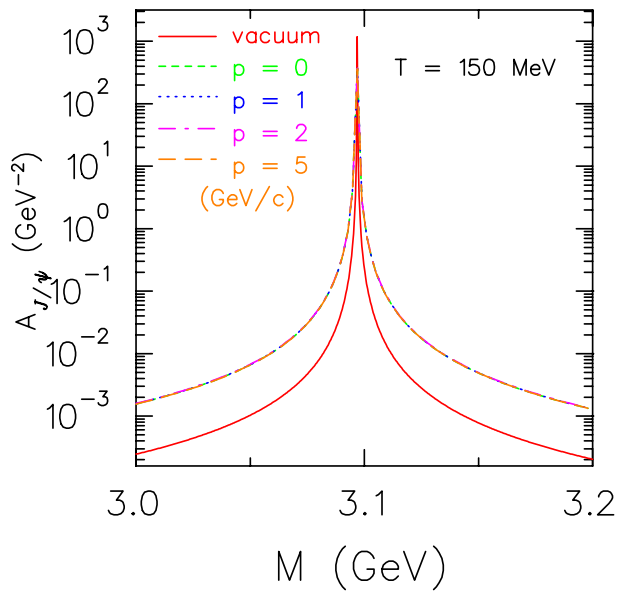}
    \caption{Spectral function in vacuum and at finite temperature 
             allowing only elastic scattering.
            }
    \label{fig5}
  \end{center}
\end{figure}

One can first shown the effects of purely elastic processes on the $J/\psi$
spectral function. This appears in Fig. \ref{fig5}. Note that 
$\Lambda$ = 2 Gev throughout this section, keeping in mind any
investigation of a more quantitative nature will need to reflect the
possible ambiguities in this choice that were discussed earlier.
Without inclusion of the elastic scattering with $\omega$'s through 
$\eta_{c}$ exchange, the spectral function deviates so little from
the vacuum that all curves lie on top of one another.  But as one
can see in Fig.\ref{fig5}, elastic scattering in the medium now 
has a non negligible effect and distorts the spectral function.  Since
the $J/\psi + \omega$ cross section with form factors is quite
flat in $\sqrt{s}$, the momentum dependence here is not important.

\begin{figure}[htbp]
  \begin{center}
  \epsfxsize 80mm 
  \epsfbox{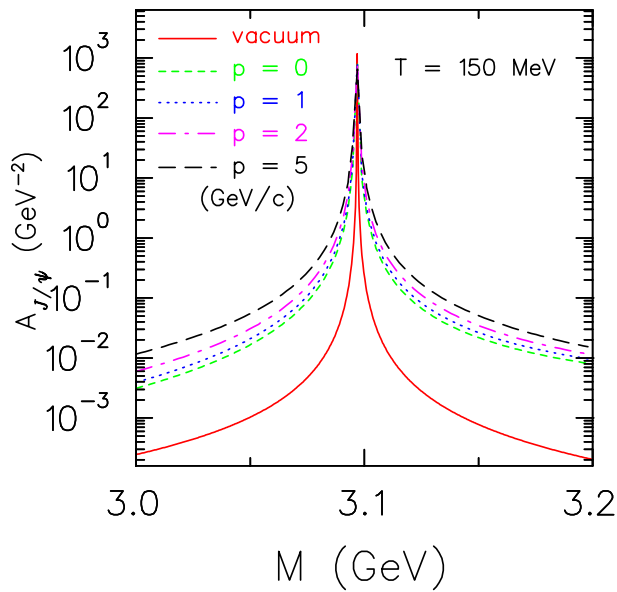}
    \caption{Spectral function in vacuum and at $T$ = 150 MeV temperature
             allowing inelastic interactions with $\pi$'s, $K$'s
             and $\rho$'s.
            }
    \label{fig6}
  \end{center}
\end{figure}

\begin{figure}[htbp]
  \begin{center}
  \epsfxsize 80mm 
  \epsfbox{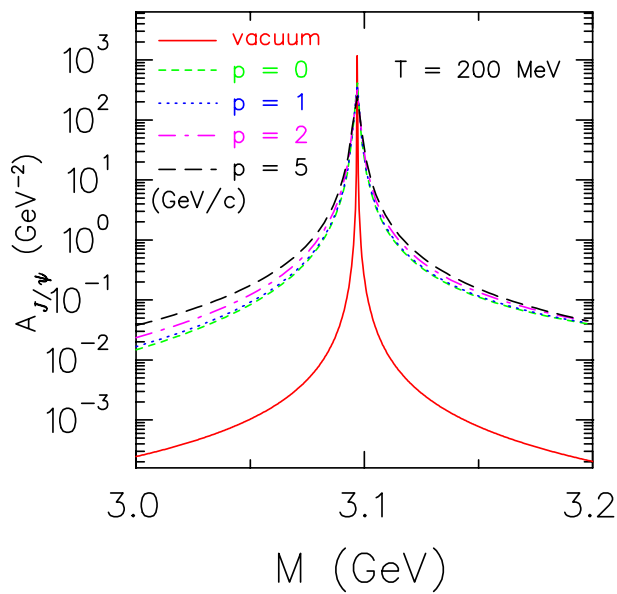}
    \caption{Spectral function in vacuum and at 200 MeV temperature
             allowing inelastic interactions with $\pi$'s, $K$'s
             and $\rho$'s.
            }
    \label{fig7}
  \end{center}
\end{figure}

Let us now consider the $J/\psi$ charmonium state traveling in a
finite temperature gas first consisting only of $\pi$'s, $K$'s and $\rho$'s. 
The spectral function at two temperatures, 150 and 200 MeV, is shown in 
Figs.~\ref{fig6} and \ref{fig7}.  One notices a substantial broadening of the
spectral distribution, along with a suppression of the peak. This considerable 
effect is even more pronounced at the higher temperature. If we include
all the inelastic processes we have considered in this work (summarized in
Table \ref{table2}), it turns out the quantitative differences between 
results involving all those  and the ones shown in 
Fig. \ref{fig6} and \ref{fig7} are small
and can be neglected. Clearly, the $\pi$'s, $K$'s, and  $\rho$'s play the
leading role. 

\section{Conclusion}

In this work,  cross sections for the interactions of
the $J/\psi$ with light mesons were evaluated.  It has been found that
those cross sections are all quite different, and also not constant with
respect to the energy of the colliding particles. 
%This suggests that
%some of the assumptions of some popular $J/\psi$ absorption models are
%over simplified. 
The form factors that are germane to meson exchange
models such as the one discussed here have been constrained by hadronic
phenomenology, Lorentz invariance and electromagnetic gauge invariance.
The numerical effect of those form factors are large and therefore a
careful treatment is mandatory. The importance of the $J/\psi + \pi \to
X$ hadronic absorption channel has been highlighted, especially those
that invole the $\eta_c$. 
Similarly, the elastic cross section involving an external $\omega$
interacting through an exchanged $\eta_c$ is found to be appreciable.
Those findings should have some effect on the hadronic phenomenology
that is an ingredient to the theoretical modeling of high energy heavy
ion collisions. Specifically, the  $J/\psi + \pi$ inelastic channel 
can be interpreted as the leading contribution in 
the $J/\psi$ absorption by 
``comovers''. In this light, the average value of  our comover 
cross section  just about reaches 2 mb. If one consider the $J/\psi +
\rho$ channel, one could even add an additional mb. A recent study 
uses $\sigma_{\rm comovers}$ = 1 mb
\cite{cape2000}.  It is important for values obtained
phenomenologically and for those based on more microscopic approaches to
eventually meet. The studies performed here however do point to the
richness of the many-body problem. Several reactions channels have been
considered and even more work needs to be done in order to complete
the survey of what turns out to be a vast hadronic landscape. Our
exploration continues.

We have  
evaluated the spectral function for a $J/\psi$ state
traveling in a finite temperature gas of mesons. We have found that the
spectral function gets considerably modified, owing mainly to inelastic
interactions with the constituents of the hot meson gas. 

For the moment, we have
refrained from attempting a detailed comparison with heavy ion data.  
It is
also clear that such an application will require
great care. For example it is claimed here that the absorption of the
$J/\psi$ on pions is important, especially when the $\eta_c$ is part of
the final state. However, inverse reactions can produce a $J/\psi$:
$\eta_c + \rho \to J/\psi + \pi$ (0.7 mb), and 
$\eta_c + \pi \to J/\psi + \rho$ (2 mb).  
The numbers in parentheses refer to cross section values that are
approximately in the centre of the form factor window.   
One therefore needs to advocate a careful simulation of the nuclear collision,
with the inclusion of the relevant important reaction channels and of
their detailed balance partners. This work is extensive, has begun, and
will be reported on elsewhere. 
It is first necessary to place
the interactions of the $J/\psi$ in a proper many-body setting, at the
appropriate level of sophistication our current understanding of hadronic
physics requires.  This also implies pointing out the caveats as well
as the successes. 

\acknowledgments
C.G. would like to thank Berndt M\"uller for a conversation during which
he made a remark that triggered the present investigation. We also
acknowledge useful discussions with T. Barnes, S. Brodsky,  D. Kharzeev, 
C. M. Ko, Z. Lin, K. Redlich, and E. S. Swanson. 
This work was
supported in part by the National Science Foundation under grant number
PHY-9814247, in part by the Natural Sciences and Engineering Research
Council of Canada, and in part by the Fonds FCAR of the Quebec
Government. 

\pagebreak
\section*{Appendix: Gauge Invariance}
\setcounter{equation}{0}
\def\theequation{A\arabic{equation}}

The amplitudes discussed in this work that couple to vector mesons that
have the quantum number of the photon need to obey gauge invariance in
the electromagnetic sector, or more specifically current conservation.
This statement is an immediate consequence of Vector Meson Dominance. 
We focus for the moment on the reaction
$J/\psi(p_{1}) + \pi(p_{2}) \to D^{*}(p_{3}) + \bar{D}(p_{4})$.  The
invariant amplitude emerging from Eq.~\ref{ellint} is 
${\cal M\/} = {\cal M\/}_{1} + {\cal M\/}_{2} + {\cal M\/}_{3}$ where
\begin{eqnarray}
{\cal M\/}_{1} & = & {\sqrt{2\over 3}}g^{2}\,\epsilon^{\mu}(p_{1})
(2p_{4}-p_{1})_{\mu}{1\over(p_{1}-p_{4})^{2}-m_{4}^{2}}
(2p_{2}-p_{3})_{\nu}\epsilon^{\nu}(p_{3})
\nonumber\\
{\cal M\/}_{2} & = & {\sqrt{2\over 3}}g^{2}\,\epsilon^{\mu}(p_{1})
\left[(2p_{3}-p_{1})_{\mu}\,g_{\nu\alpha} \, - \, 
(p_{1}+p_{3})_{\alpha}\,g_{\mu\nu} \, + \, (2p_{1}-p_{3})_{\nu}\,g_{\mu\alpha}
\right]
\nonumber\\
& \ & \times{\left[-g^{\alpha\beta} + (p_{1}-p_{3})^{\alpha}
(p_{1}-p_{3})^{\beta}/m_{3}^{2}\right]\over(p_{1}-p_{3})^{2}-m_{3}^{2})}
(p_{2}+p_{4})_{\beta}\epsilon^{\nu}(p_{3})
\nonumber\\
{\cal M\/}_{3} & = & {\sqrt{2\over 3}}g^{2}\,\epsilon^{\mu}(p_{1})
\left[-g_{\mu\nu}\right]\epsilon^{\nu}(p_{3}).
\end{eqnarray}

Gauge invariance requires that ${\cal{M}\,}\left(\epsilon^{\mu}(p_{1}) \to 
p_{1}^{\mu}\right)$ must be identically zero.
In particular, it must vanish for arbitrary choices of pseudoscalar and
vector masses. 

Replacing $\epsilon^{\mu}(p_{1})\to p_{1}^{\mu}$, and doing the 
contraction gives
\begin{eqnarray}
{\cal M\/}_{1} & = & {\sqrt{2\over 3}}g^{2}\,
\left[-(2p_{2}-p_{3})_{\nu}\right]\,\epsilon^{\nu}(p_{3})
\nonumber\\
{\cal M\/}_{2} & = & {\sqrt{2\over 3}}g^{2}\,\left\lbrace(p_{2}+p_{4})_{\nu}
+ {1\over(p_{1}-p_{3})^{2}-m_{3}^{2}}\left[
{ (2p_{3}-p_{1})\cdot p_{1}(p_{1}-p_{3})\cdot(p_{2}+p_{4})(p_{1}-p_{3})_{\nu}
\over m_{3}^{2}}
\right. \right.
\nonumber\\
&\ & + (p_{1}+p_{3})\cdot(p_{2}+p_{4})(p_{1})_{\nu}
-{(p_{1}+p_{3})\cdot (p_{1}-p_{3})(p_{1}-p_{3})\cdot(p_{2}+p_{4})(p_{1})_{\nu}
\over m_{3}^{2}}
\nonumber\\  
&\ & \left. \left. - (2p_{1}-p_{3})_{\nu}(p_{1})\cdot(p_{2}+p_{4})
+{(2p_{1}-p_{3})_{\nu} (p_{1})\cdot(p_{1}-p_{3})(p_{1}-p_{3})\cdot(p_{2}
+p_{4})
\over m_{3}^{2}} \right] \right\rbrace\,\epsilon^{\nu}(p_{3})
\nonumber\\
{\cal M\/}_{3} & = & {\sqrt{2\over 3}}g^{2}\,\left[
- (p_{1})_{\nu}\right]\epsilon^{\nu}(p_{3}).
\end{eqnarray}

We note that ${\cal M\/}_{1}$ plus the first term in ${\cal M\/}_{2}$,
plus ${\cal M\/}_{3}$ vanishes due to energy-momentum conservation.
The remaining pieces from ${\cal M\/}_{2}$ are
\begin{eqnarray}
& \propto & {1\over m_{3}^{2}\left[(p_{1}-p_{3})^{2}-m_{3}^{2}\right]}
{\Huge\lbrace} 
(2p_{3}-p_{1})\cdotp_{1}(p_{1}-p_{3})\cdot(p_{2}+p_{4})(p_{1}-p_{3})_{\nu}
+ m_{3}^{2}(p_{1}+p_{3})\cdot(p_{2}+p_{4})(p_{1})_{\nu}
\nonumber\\
& \ &
-(p_{1}+p_{3})\cdot(p_{1}-p_{3})(p_{1}-p_{3})\cdot(p_{2}+p_{4})(p_{1})_{\nu}
- m_{3}^{2}(2p_{1}-p_{3})_{\nu}p_{1}\cdot(p_{2}+p_{4})
\nonumber\\
& \ & 
+(2p_{1}-p_{3})_{\nu}p_{1}\cdot(p_{1}-p_{3})(p_{1}-p_{3})\cdot(p_{2}
+p_{4})  {\huge\rbrace}\,\epsilon^{\nu}(p_{3})
\end{eqnarray}

Terms proportional to $(p_{3})_{\nu}$ vanish when contracted with
$\epsilon^{\nu}(p_{3})$ due to transversality.   Thus, the surviving 
terms can be
written as
\begin{eqnarray}
& \propto & {(p_{1})_{\nu}\over 
m_{3}^{2}\left[(p_{1}-p_{3})^{2}-m_{3}^{2}\right]}
{\huge\lbrace} 
(2p_{3}-p_{1})\cdot(p_{1})(p_{4}-p_{2})\cdot(p_{2}+p_{4})
+ m_{3}^{2}(p_{1}+p_{3})\cdot(p_{2}+p_{4})
\nonumber\\
& \ & 
+ m_{3}^{2}(-2p_{1})\cdot(p_{2}+p_{4})
-(p_{1}+p_{3})\cdot(p_{1}-p_{3})(p_{4}-p_{2})\cdot(p_{2}+p_{4})
+2(p_{1})\cdot(p_{1}-p_{3})(p_{4}-p_{2})\cdot(p_{2}+p_{4})
{\huge\rbrace}\,\epsilon^{\nu}
\end{eqnarray}

Finally, we can simplify to
\begin{eqnarray}
 & \propto & (p_{1})_{\nu}{\huge\lbrace}
(m_{4}^{2} - m_{2}^{2})\left[ m_{3}^{2} \right]
+ m_{3}^{2} \left[ (p_{3}-p_{1})\cdot(p_{2}+p_{4})\right]
{\huge\rbrace}\,\epsilon^{\nu}(p_{3}) 
\nonumber\\
 & = & (p_{1})_{\nu}{\huge\lbrace}
m_{3}^{2}(m_{4}^{2}-m_{2}^{2})
+ m_{3}^{2}(p_{2}-p_{4})\cdot(p_{2}+p_{4})
{\huge\rbrace}\,\epsilon^{\nu}(p_{3})
\nonumber\\
& = & (p_{1})_{\nu} {\huge\lbrace} m_{3}^{2}{\huge\lbrace}
m_{4}^{2} - m_{2}^{2} + m_{2}^{2} - m_{4}^{2}{\huge\rbrace}
\nonumber\\
& = & 0.
\end{eqnarray}

Indeed, current is conserved in the general case.
The other channels can similarly be shown to conserve current.

%   the references....

%\begin{figure}[b!]  
%\end{figure}
\newpage

\begin{table}
\caption{Model prediction for widths. The phenomenological approach
refers to that of Refs.~\protect\cite{jpsixsect2,kevin,linko}.}
\label{table1}
\begin{tabular}{c|c|c|c}
particle
& chiral model &pheno. model & experiment \\
\tableline
K$^*$(892)$^{\rm 0}$ & 44.5 MeV & 97.0 MeV&50.5 $\pm$ 0.6 MeV \\ \hline
K$^*$(892)$^{\pm}$ & 44.5 MeV & 97.0 MeV & 49.8 $\pm$ 0.8 MeV \\ \hline
D$^*$(2007)$^{\rm 0}$ & 10.1 keV & 22.0 KeV &$<$ 2.1 MeV, 90\% CL\\ \hline
D$^*$(2010)$^{\pm}$ & 21.1 keV & 46.0 KeV & $<$ 131 keV, 90\% CL\\ 
\end{tabular}
\end{table}

\begin{table}
\caption{We list here the hadronic reactions involving $J/\psi$ that were
considered in this work. It is implied that the figures also include
the Hermitian conjugate inelastic final state when it is different 
from the one listed below.}
\label{table2}
\begin{tabular}{lcll}
\multicolumn{1}{c}{Elastic channels} & & \multicolumn{2}{c}{Inelastic
channels}\\
      &          &  initial state &final state\\
\tableline
$J/\psi + \pi $  &  & $J/\psi + \pi$ & $D^* + \bar{D}$\\
$J/\psi + \eta$  &  & $J/\psi + \pi$ &$\eta_c  + \rho$\\
$J/\psi + K$     &  & $J/\psi + \pi$ &$\eta_c  + b_1 $\\
$J/\psi + \rho$  &  & $J/\psi + \eta$ &$D^* + \bar{D}$\\  
$J/\psi + \omega$&  & $J/\psi + \eta$ &$\eta_c + \phi$\\
$J/\psi + \phi $ &  & $J/\psi + K   $ & $D_s + \bar{D^*}$\\
                 &  & $J/\psi + \rho$ & $D + \bar{D}$\\    
                 &  & $J/\psi + \rho$ & $D^* + \bar{D^*}$\\
                 &  & $J/\psi + \rho$ & $\eta_c + \pi$\\  
                 &  & $J/\psi + \omega$&$D + \bar{D}$\\
                 &  & $J/\psi + \omega$&$D^* + \bar{D^*}$\\
                 &  & $J/\psi + \phi $ &$D + \bar{D}$\\
                 &  & $J/\psi + \phi $ &$D^* + \bar{D^*}$\\
                 &  & $J/\psi + K    $ & $\eta_c + K_1$\\
                 &  & $J/\psi + K^*  $ &$D_s + \bar{D}$\\
                 &  & $J/\psi + K^*  $ &$D_s^* + \bar{D^*}$\\
\end{tabular}
\end{table}


\begin{references}
\bibitem{expln} We examine here some claim made in the recent
literature \protect\cite{linko}. The Lagrangian used here
(Eq.~(\protect\ref{ellint})) is formally equivalent to that in Eq. (6) of
Ref.~\protect\cite{linko}, the reader can be convinced by
inspection. With the ``phenomenological'' version of our approach (as
explained in the text), our total $J/\psi + \pi$ and $J/\psi + \rho$
cross sections (our Fig.~\protect\ref{fig4b}) are identical to those in
Fig.~2 of Ref.~\protect\cite{linko}. 
As correctly pointed out in the latter reference,
Ref.~\protect\cite{kevin} does contain typos that make an 
identification with the approach used here and in
Ref.~\protect\cite{linko} difficult. However, those typos have not
propagated to the calculations. What was calculated in
Ref.~\protect\cite{kevin} for the $J/\psi + \pi$ cross section was a
partial cross section, for a given charged state of the participating
fields. There are four such channels. The $J/\psi + \pi$ result of
Ref.~\protect\cite{kevin} multiplied by four is equivalent to the one
obtained here and to that of \protect\cite{linko}.
\end{references}
\end{document}